\documentclass[twocolumn,amsmath,amssymb]{revtex4}   	
\usepackage{graphicx}				
\usepackage{color}		
\usepackage{pdfpages}

\begin{document}
\title{Scaling of the Non-Phononic Spectrum of Two-Dimensional Glasses}

\date{\today}

\author{Lijin Wang$^{1}$}
\author{Grzegorz Szamel$^{2}$}
\author{Elijah Flenner$^{2}$}

\affiliation{$^1$School of Physics and Optoelectronic Engineering, Anhui University, Hefei 230601, China}

\affiliation{$^2$Department of Chemistry, Colorado State University, Fort Collins, Colorado 80523, USA}

\begin{abstract}
Low-frequency vibrational harmonic modes of glasses are frequently used to understand their
universal low-temperature properties. One well studied feature is the excess low-frequency density of states over
the Debye model prediction.
Here we examine the system size dependence of the  density of states for two-dimensional glasses.
For systems of fewer than 100 particles, the  density of states scales with the system size as if all the modes were plane-wave-like.
However, for systems greater than 100 particles we find a different system-size scaling of the cumulative density of states below the first
transverse sound mode frequency, which can be derived from the assumption that these modes are quasi-localized. Moreover, for systems
greater than 100 particles, we find  that the cumulative
density of states scales with frequency as a power law with the  exponent that leads to the exponent $\beta=3.5$ for the density of states independent of system size.
\end{abstract}

\maketitle

A striking feature of glasses is the excess of low-frequency vibrational harmonic modes over that predicted by
Debye theory. Many computational studies analyzed low-frequency vibrational modes
in glasses with the goal to characterize the excess harmonic modes, which are
frequently referred to as non-phononic modes. Since the same low-temperature properties are found in different glasses,
one is naturally tempted to look for universal features of the excess modes, independent of the model and the glass formation protocol.
Importantly, these features should hold in the thermodynamic limit. In many studies
the implicit hypothesis is that the excess modes density $\mathcal{D}_{ex}(\omega)$ follows a power law,
$\mathcal{D}_{ex}(\omega) = A_{ex} \omega^{\beta}$.

Two methods are generally employed to determine $\beta$. One approach
is to separate modes by
their participation ratio with the assumption that the low-frequency excess modes have a significantly
smaller participation ratio than the extended, plane-wave-like modes \cite{Schober1991,Wang2019nc,Mizuno2017}. This method was
used to argue that $\beta = 4$ in three dimensions \cite{Wang2019nc,Mizuno2017}.
However, this method fails in two dimensions since it is impossible to
choose a suitable participation ratio cutoff to differentiate between the excess modes and the plane-wave-like modes in two dimensions.

The second method
is to study small systems \cite{Lerner2016,Kapteijns2018,Lerner2020}.
The frequency range of the first band of plane-wave-like modes scales as $L^{-1}$ where $L$ is
the simulation box size. The assumption is that the modes below this band of plane-wave-modes represent the excess modes
and one can determine the scaling of $\mathcal{D}_{ex}(\omega)$ from studying the harmonic mode spectrum for frequencies
significantly below those of the first band. While conceptually
straight forward, there are some subtle issues that must be kept in mind. First, every
simulated glass has a slightly different shear modulus that results in a slightly different frequency expected for the first plane-wave-like
mode, $\omega_{T1}$, for an ideal elastic body. In addition, the modes are not pure plane waves and there is a distribution of frequencies
around $\omega_{T1}$ for each glass. For  these reasons one can have a rather large range of frequencies where the first plane-wave-like
modes exist. If one does not separate the plane-wave-like modes (like in the first method),
one has to be careful to infer $\mathcal{D}_{ex}(\omega)$ using only modes that do not correspond to plane-wave-like modes.

Another difficulty is that there is a reported finite size effect influencing the exponent $\beta$ in three-dimensional glasses~\cite{Lerner2020}.
It has been reported that $\beta < 4$ for
small systems and $\beta$ equals 4 for large enough systems \cite{Lerner2020}. Recall that the plane-wave-like modes move to
lower frequencies with increasing system size, and thus there is 
a smaller frequency range where
the excess modes spectrum is found. In addition, in two dimensions there are very few excess modes~\cite{Mizuno2017,Wang2021} and a very large number
of glass realizations are needed to obtain good enough statistics to determine an accurate value for $\beta$.
Analysis of $\beta$ in  two dimensions is complicated by a reported finite size effect in the pre-factor $A_{ex}$, $A_{ex} \sim \left(\log N\right)^{(\beta+1)/2}$  with $N$
the number of particles in the system \cite{Kapteijns2018}.

In a previous publication we stated that there was no clear evidence that there is a finite size effect for $\beta$
in simulations of two-dimensional glasses \cite{Wang2021}. In contrast,
Lerner and Bouchbinder argued very recently \cite{Lerner2022} that $\beta$ increases with increasing system
size and is four for a large enough (N=1600) aged glass.

Here we present analysis of our results of
the cumulative density of states $I(\omega) = \int_{0^+}^\omega D(\omega^\prime) d\omega^\prime$, where $D(\omega)$ is
the full density of states, that lead to our conclusions that there is no system size dependence of $\beta$ and
examine in detail some finite size effects of the density of states observed in two-dimensional glasses.
We also argue 
that there is no clear evidence of systematic system size dependence observed in the data presented Ref.~\cite{Lerner2022}.

\begin{figure}
\includegraphics[width=0.9\columnwidth]{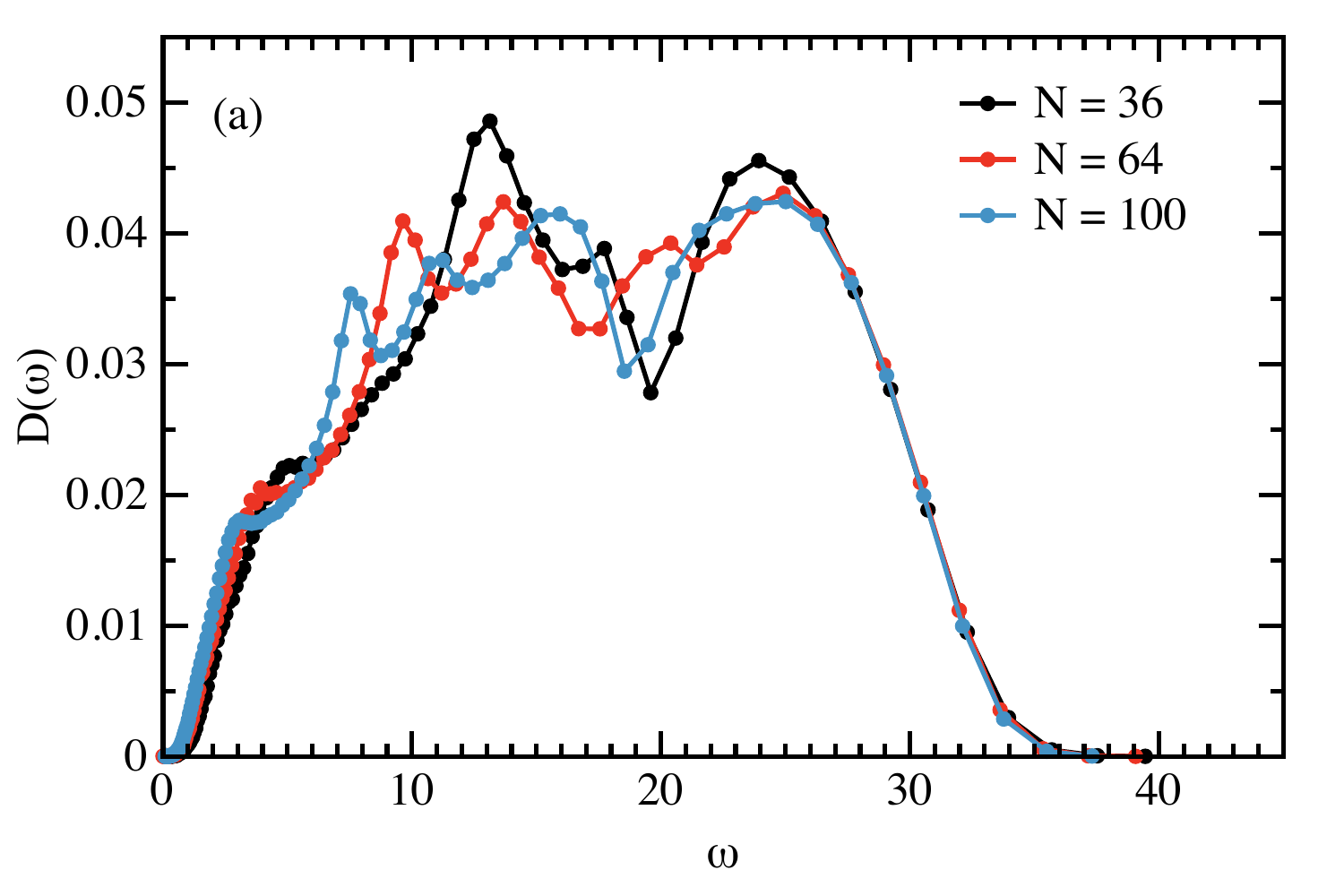}
\includegraphics[width=0.9\columnwidth]{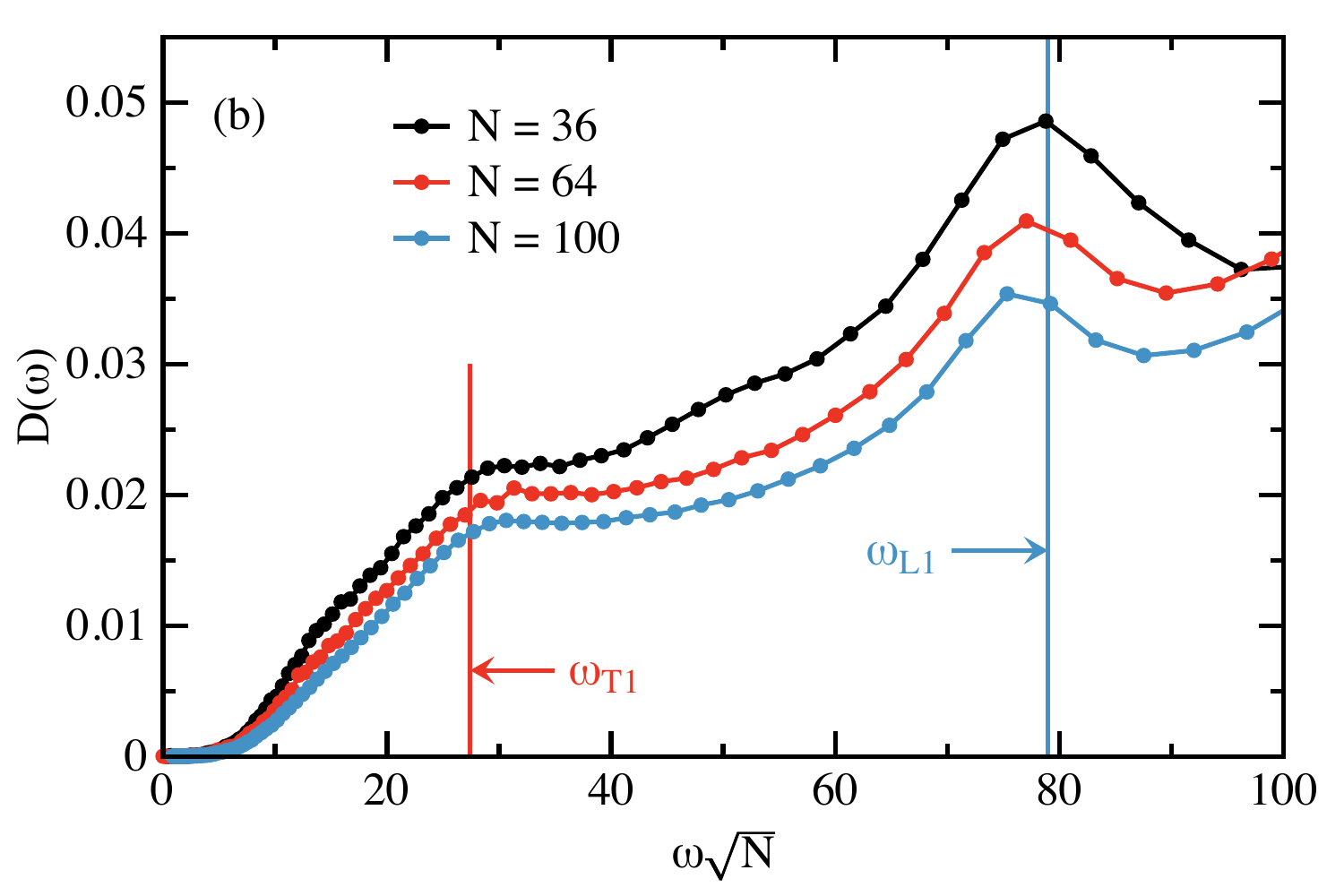}
\caption{\label{dos}(a) The density of states of small two-dimensional glasses with $N$ paticles. There is a distinct system size dependence with
a shoulder at low frequencies and peaks showing up at $N$ dependent frequencies. (b) The density of states versus $\omega \sqrt{N}$. The shoulder
is shown to correspond to $\omega_{T1} = \sqrt{G/\rho} (2 \pi/L)$ and the first peak corresponds to $\omega_{L1} = \sqrt{(G+K)/\rho} (2 \pi/L)$. }
\end{figure}

For an ideal elastic body, the  frequency of  the first transverse wave
is given by $\omega_{T1} = \sqrt{(G/\rho)} (2\pi/L)$, where $G$ is the
shear modulus, $\rho$ is the mass density, and $L$ is the linear size of the sample.
The frequency of the first longitudinal wave is
given by $\omega_{L1} = \sqrt{(G+K)/\rho} (2\pi/L)$, where $K$ is the bulk modulus.
We start our analysis of the IPL-10 system discussed in Ref.~\cite{Wang2021} for systems
of $N \le 100$. Shown in Fig.~\ref{dos}
is the density of states $D(\omega) = (2N-2)^{-1} \sum_n \delta(\omega-\omega_n)$ for $N=36$, 64, and 100.
The frequencies $\omega_n$ are obtained by diagonalizing the Hessian matrix.
For these system sizes
there are two distinct features at low frequencies, a shoulder and a peak. In Fig.~\ref{dos}(b) we scale the frequency by $\sqrt{N} \sim L$
and find that the two low frequency features align. The vertical lines are the frequencies of ideal plane waves, $\omega_{T1}$ (red) and
$\omega_{L1}$ (blue). Fig.~\ref{dos}(b) suggests that the shoulder is associated with the first transverse wave and the peak is associated
with the first longitudinal wave.

To determine the scaling exponent of the density of states, $\beta$, we examine the cumulative density of states $I(\omega)$.
Unlike the density of states, the cumulative density of states does not depend on the binning algorithm. 
The frequency dependence of the cumulative density of states  follows from the scaling of $D(\omega^\prime)$ for values of
$\omega^\prime$ less than $\omega$. Since the hypothesis is that the lowest frequencies below $\omega_{T1}$ represent $\mathcal{D}_{ex}(\omega)$,
we should be able to infer $\mathcal{D}_{ex}(\omega)$ from the low frequency behavior of $I(\omega)$.

\begin{figure}
\includegraphics[width=0.9\columnwidth]{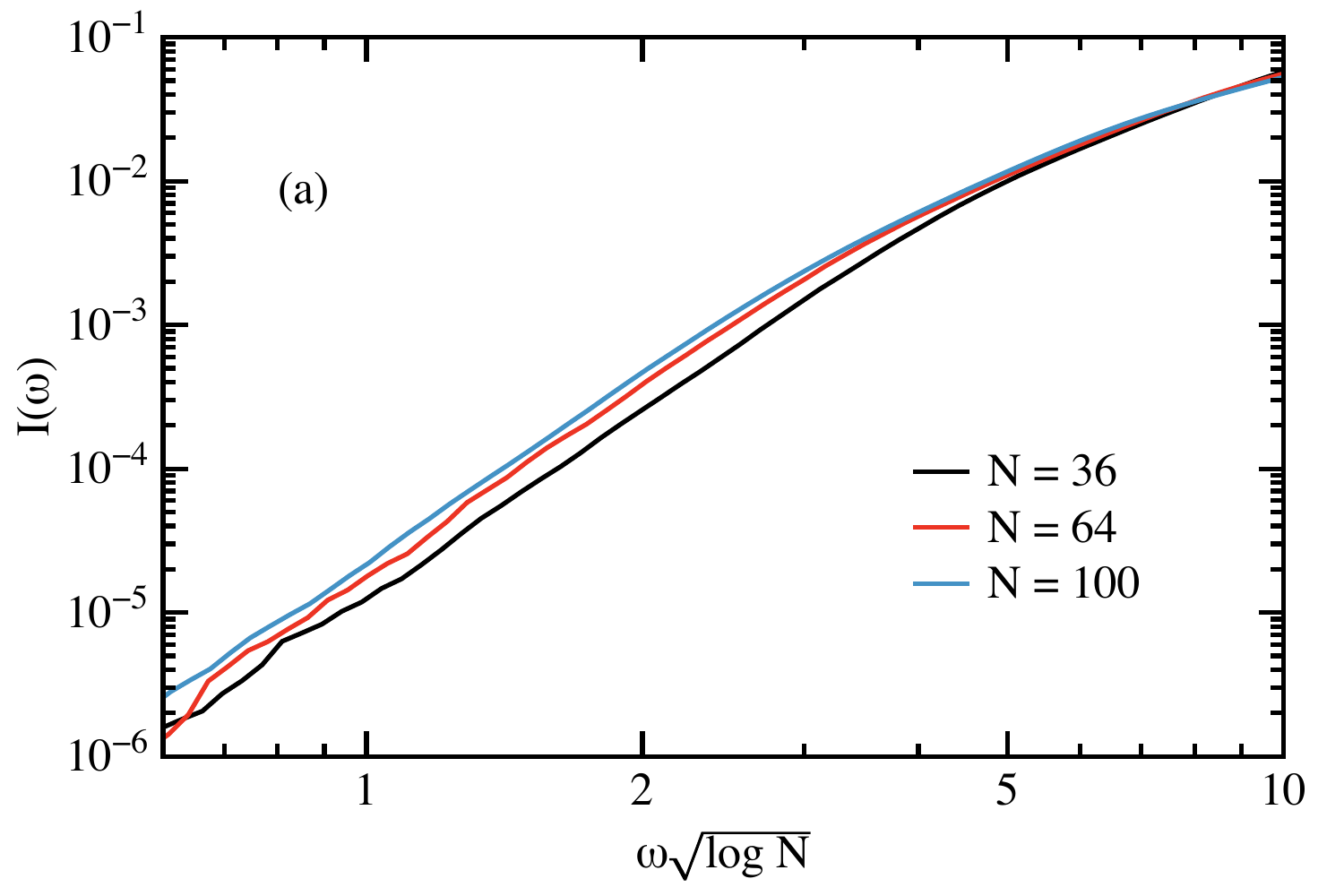}
\includegraphics[width=0.9\columnwidth]{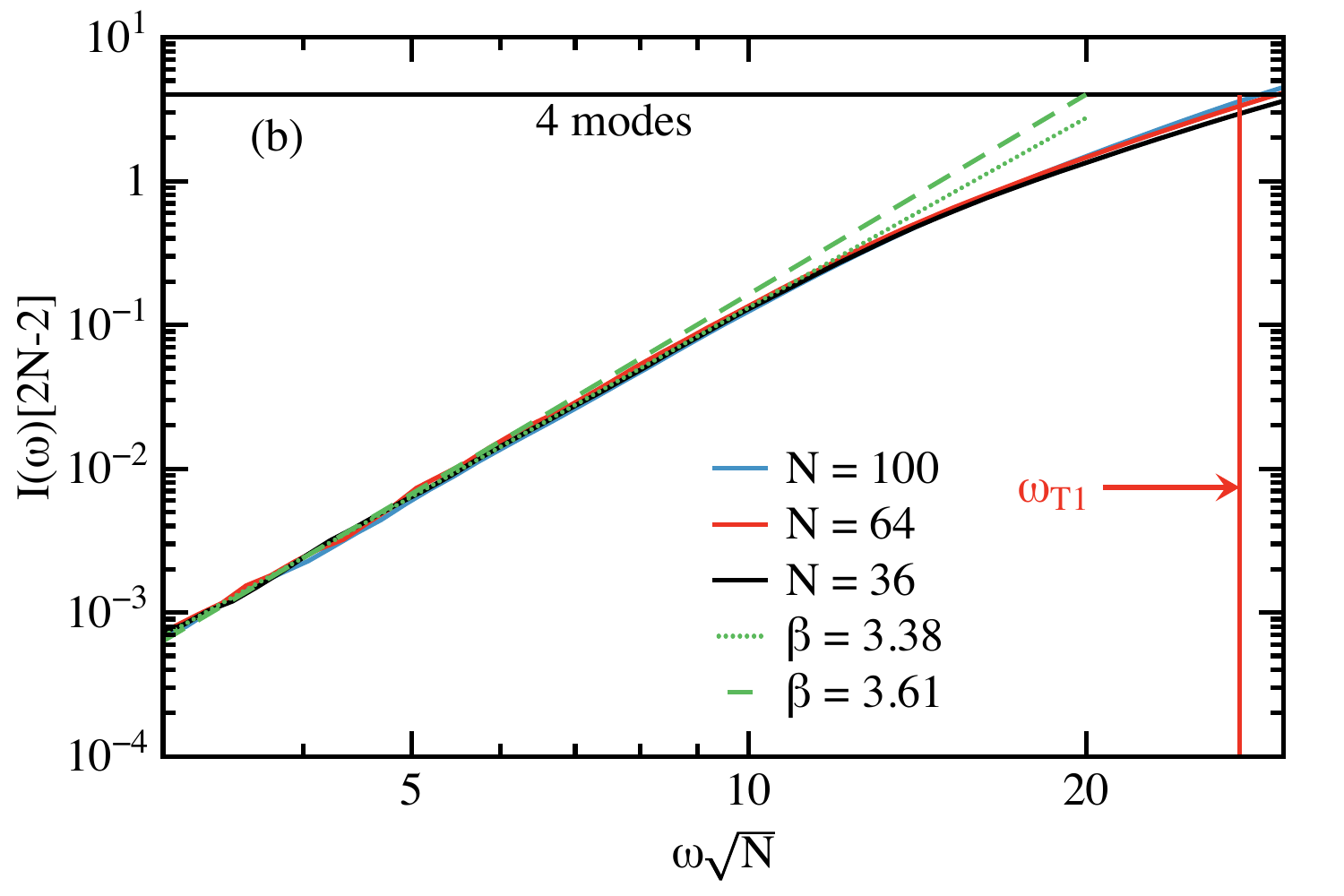}
\caption{\label{small}(a) Scaling of the cumulative density of states proposed in Ref.~\cite{Lerner2018}. There are
small deviations from this scaling for these small system sizes. (b) Expected scaling of the cumulative density of
states if all the modes are plane-wave-like. This scaling results in better data collapse than the scaling
proposed in Ref.~\cite{Lerner2018} for these small system sizes. The red vertical line is the frequency
$\omega_{T1}$ expected for an ideal transverse wave. For an ideal amorphous elastic body,
$I(\omega)(2N-2) = 4$ at $\omega_{T1}$. The dotted green line is a fit to the $N=36$ particle results and
the dashed green line is a fit to the $N=64$ particle results; the corresponding   $\beta$ in the
fitting function $I(\omega) [2N-2] \sim (\omega \sqrt{N})^{\beta +1}$ is indicated in the legend. Both fits are from $3 \le \omega \sqrt{N} \le 6$. }
\end{figure}

\begin{figure}
\includegraphics[width=0.9\columnwidth]{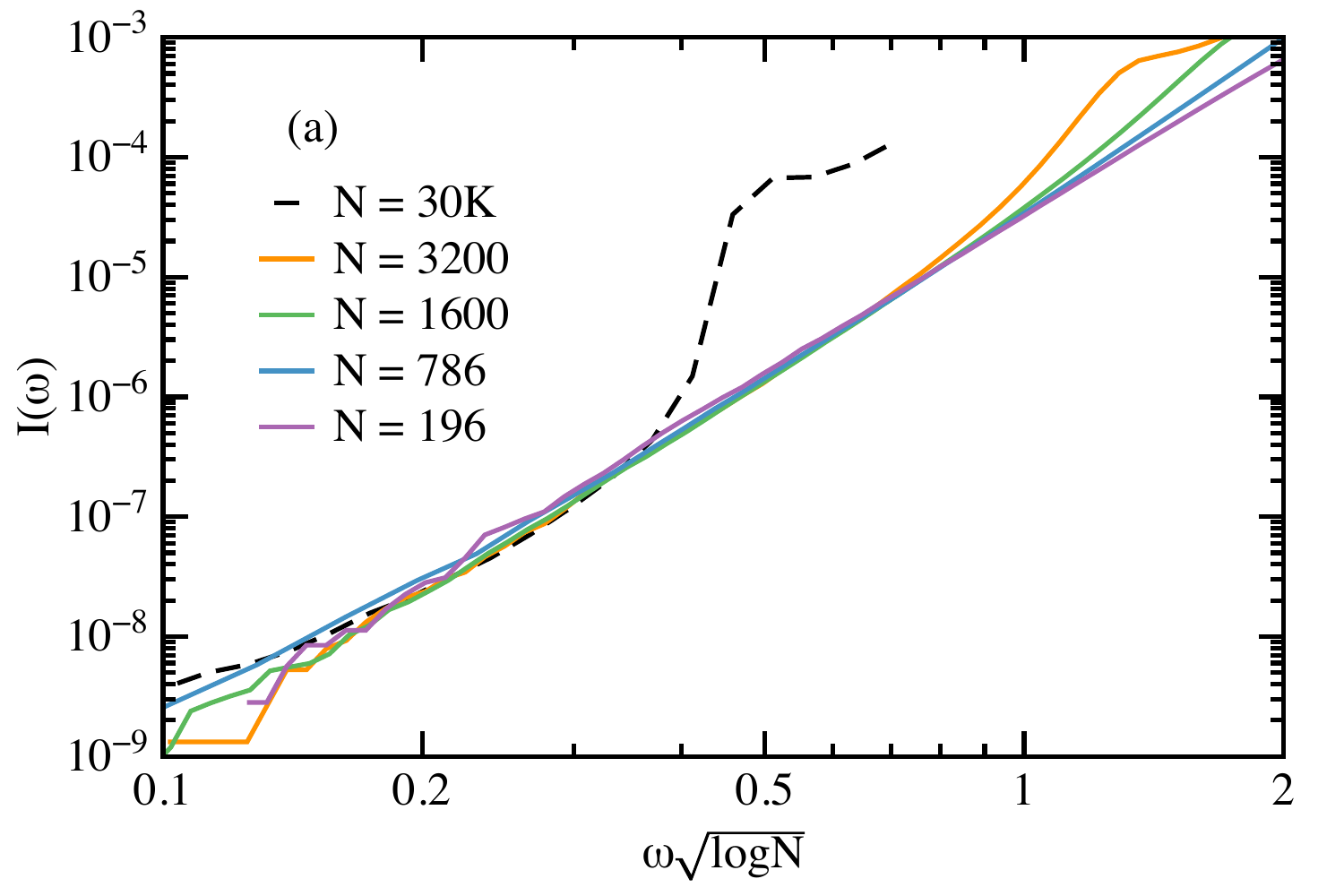}
\includegraphics[width=0.9\columnwidth]{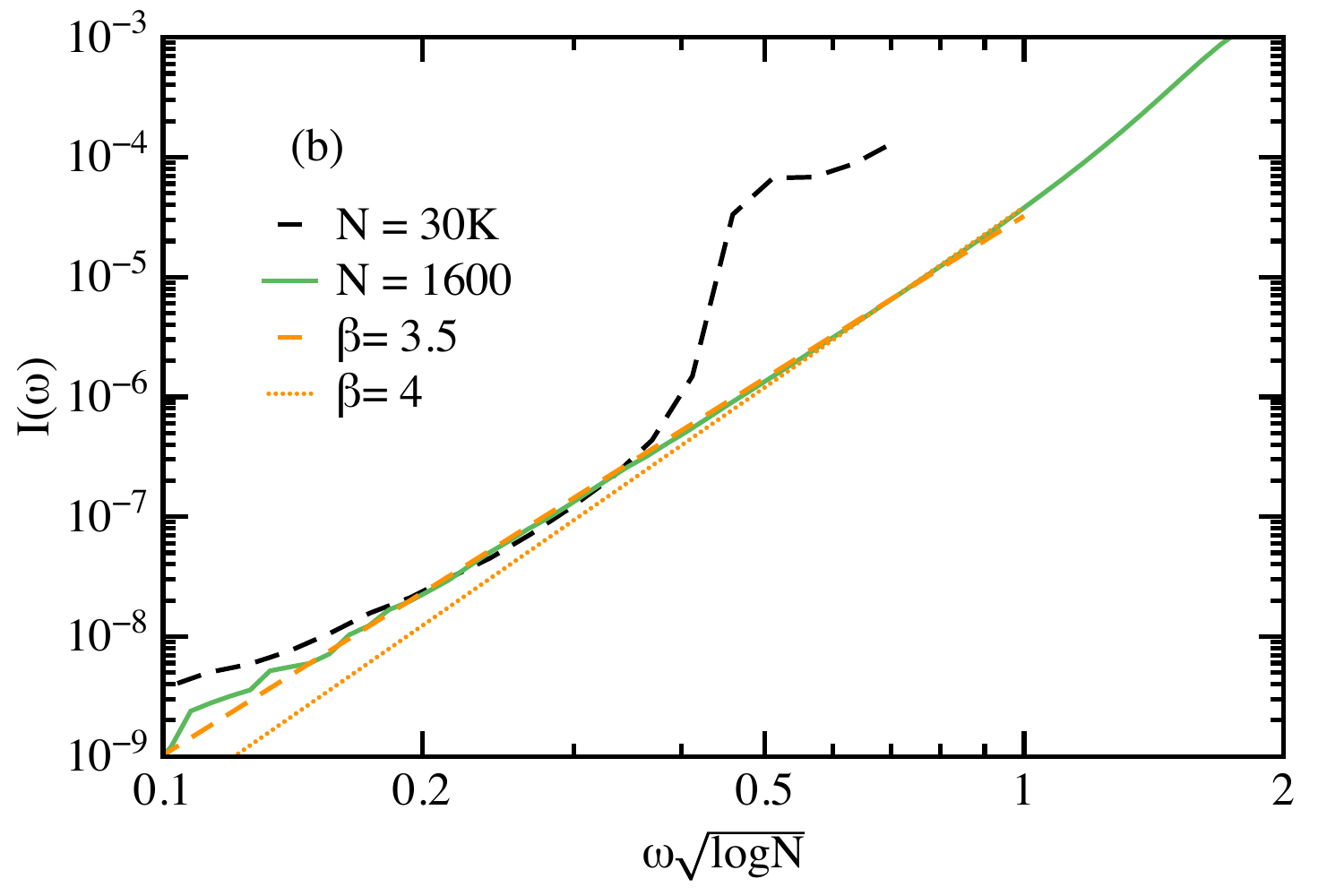}
\caption{\label{large}(a) Scaling plot of the cumulative density of states for system sizes $196 \le N \le 30\, 000$ for the IPL-10 system~\cite{Wang2021}. Unlike for smaller
systems, the $I(\omega) = f(\omega \sqrt{\log N})$ scaling produces reasonable data collapse at small frequencies. There is some
deviation at low frequencies, but this cannot be statistically ruled out. (b) Direct comparison for
systems of $N = 1600$ and $N = 30\, 000$. We fit the $N = 1600$ particle system from $0.2 \le \omega \sqrt{\log N} \le 0.8$
to $I(\omega \sqrt{\log N}) \sim (\omega\sqrt{\log N})^{\beta +1}$.
The dotted orange line corresponds to a fit with $\beta = 4$ fixed,
while the dashed orange line corresponds to a fit with $\beta = 3.5$ fixed.  }
\end{figure}

We examined the scaling of $I(\omega)$ for very small systems of $N\le 100$ particles. Shown in Fig.~\ref{small}(a) is
$I(\omega)$ versus $\omega \sqrt{\log N}$, which is a scaling predicted
from continuum elasticity due to a $r^{-1}$ decay of the non-phononic modes at large distances $r$ \cite{Kapteijns2018,Lerner2018}.
There is a systematic deviation from this scaling. In contrast, we
find nearly perfect data collapse of $I(\omega) [2N-2]$ versus $\omega  \sqrt{N}$, Fig.~\ref{small}(b), which is a scaling
motivated by the system size dependence of Debye theory. We are  counting the average number of modes up to $\omega \sqrt{N}$ and
seeing how many fall below $\omega_{T1}$.  For an ideal elastic body, $I(\omega)[2N-2] = 4$ at
$\omega_{T1} \sqrt{N}$ (vertical red line). For these very small systems
we do not find any evidence of excess modes and do not believe that the excess mode scaling can be inferred from
systems less than 100 particles. This analysis also highlights the difficulty in determining the frequency range
to use to study excess (non-phononic) modes.

We fit $I(\omega) [2N-2]$ versus $\omega \sqrt{N}$ for $3 \le \omega \sqrt{N} \le 6$ for $N = 36$, 64 and 100
to $a (\omega \sqrt{N})^{\beta +1}$. We find that $\beta$ varies from $3.38$ for $N=36$ to $3.61$ for $N=64$, and
that $\beta  = 3.48$ for $N = 100$. We show the fits for $N=36$ and $N=64$ in Fig.~\ref{small}(b). It is impossible to
determine from the fits if one value of $\beta$ should be preferred over another. Ideally, we would want to greatly expand
the fitting range to claim a specific power law, but it is very difficult to find a large range of frequencies in which a power law is obeyed.

We note that the $I(\omega) = f(\omega\sqrt{\log N})$ scaling is obeyed for $100 < N \le 30\, 000$, Fig.~\ref{large}(a).
Since the $f(\omega \sqrt{\log N})$ scaling is derived from continuum elasticity \cite{Lerner2018,Kapteijns2018}, it is unsurprising that
this scaling breaks down for small systems. There are some deviations from scaling at the smallest frequencies,
but we believe that better statistics would
improve the overlap. In Fig.~\ref{large}(b) we show only the $N = 1600$ and $N = 30\, 000$ results. We fit the $N = 1600$ results for
$0.2 \le \omega \sqrt{\log N} \le 0.8$ to $I(\omega \sqrt{\log{N}}) \sim (\omega \sqrt{log N})^{\beta +1}$
where we fixed $\beta = 4$ (orange dotted line) and $\beta = 3.5$ (orange dashed line).
The $\beta = 3.5$ fit provides a better description of the $N = 30\, 000$ results. Note that we cannot statistically rule out
$\beta = 4$ on this analysis alone, but we will show that $\beta \approx 3.5$ is consistent with all our results for systems over $100$ particles.

\begin{figure}
\includegraphics[width=0.9\columnwidth]{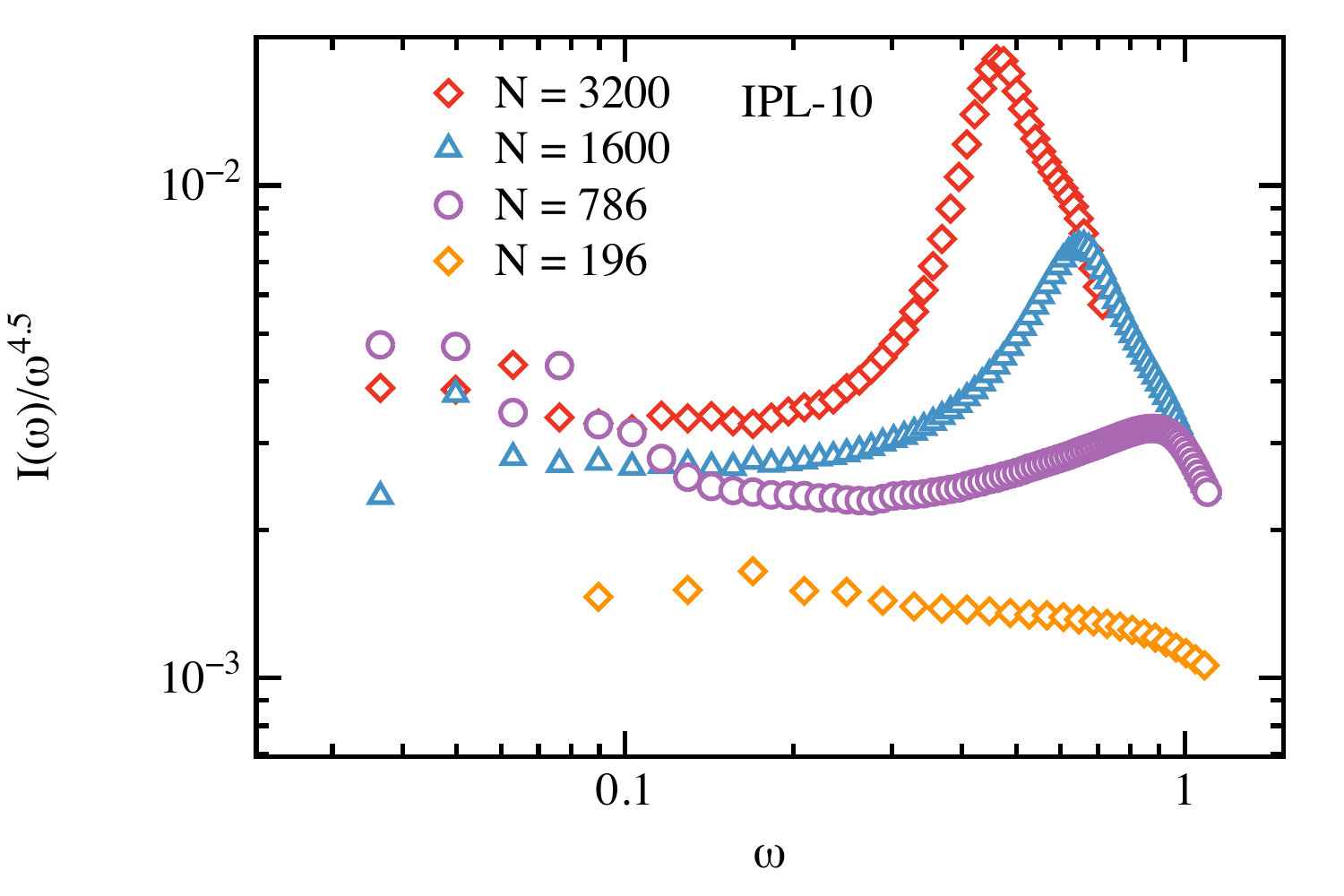}
\caption{\label{ipl10}The cumulative density of states for the IPL-10 system for different system sizes studied in Ref.~\cite{Wang2021}. Note that this figure is a reproduction of  Fig. 2(b) of  the supplemental material of  Ref.~\cite{Wang2021}.
}
\end{figure}

Shown in Fig.~\ref{ipl10} is $I(\omega)/\omega^{4.5}$ (corresponding to $\beta = 3.5$)
for the IPL-10 system for different system sizes discussed in Ref.~\cite{Wang2021}.
The low frequency data suggests that $\beta \approx 3.5$ for every system size except for $N = 786$.
As pointed out in Ref.~\cite{Lerner2022}, the $N=786$ particle result shows an upturn at small
frequencies suggesting a $\beta$ different than the other system sizes.
We find that this system size for the IPL-10 system is an outlier and it does not provide
sufficient evidence of a finite size effect.
As we show below in Fig.~\ref{N786}, with improved statistics
we find that $\beta$ for $N=786$ is consistent with our results for
other systems and other system sizes for the same system.

\begin{figure}
\includegraphics[width=0.9\columnwidth]{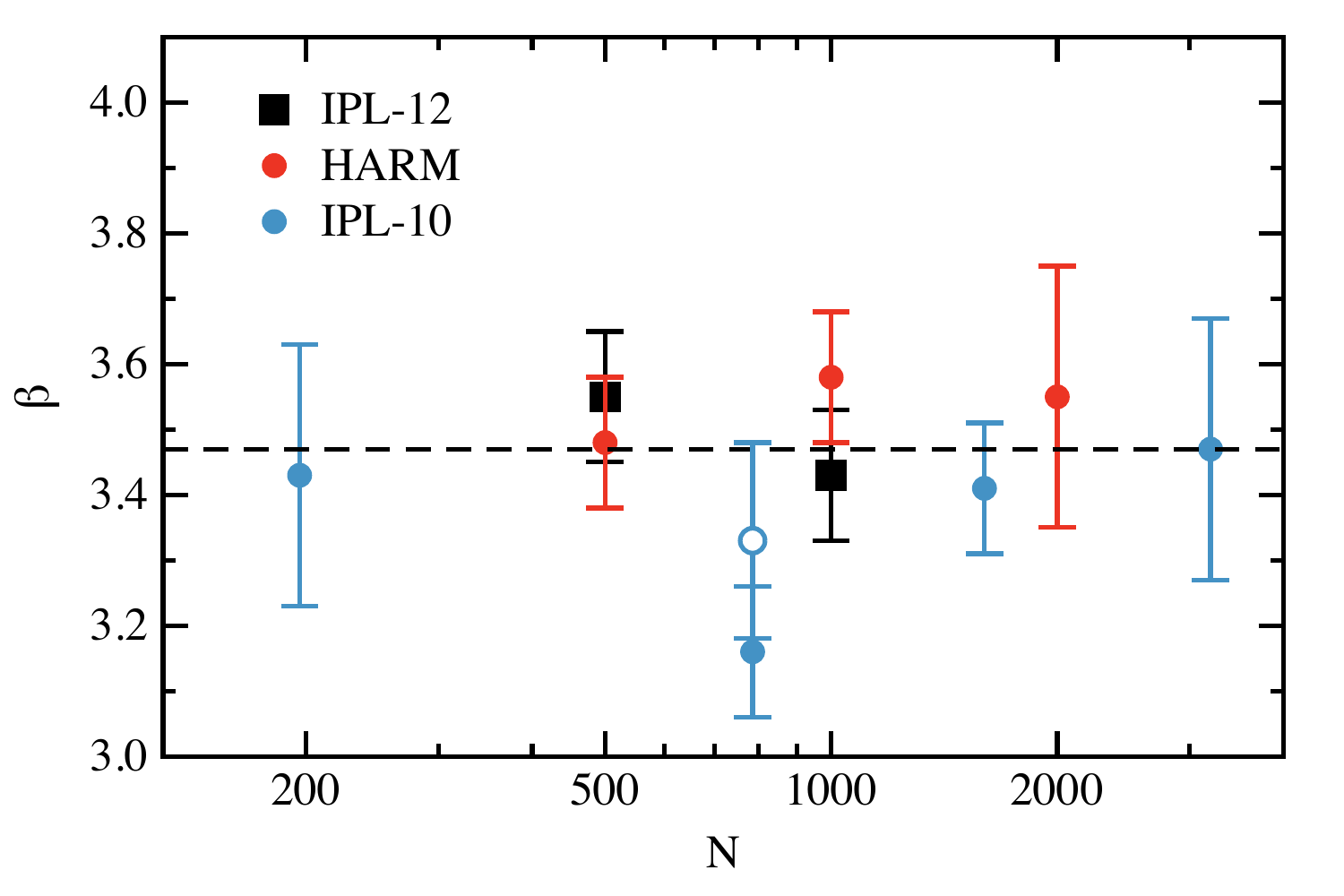}
\caption{\label{Lijin}The exponent $\beta$ obtained from fits of the cumulative density of states
$I(\omega)$ for the systems studied by Wang \textit{et al.}\ in Ref. \cite{Wang2021}. The horizontal line is the average
value of $\beta$.}
\end{figure}

To determine the exponent $\beta$ for systems of more than 100 particles
we fit $\log [I(\omega)] = (\beta+1)\log (\omega)+a$. For each system size we fit three to four frequency ranges to obtain
$\beta$ and average $\beta$ obtained from these fits. The uncertainty is
chosen to include each value of $\beta$ for a given system and system size. We exclude the results for the Lennard-Jones systems presented in
Ref. \cite{Wang2021} since there is a systematic downturn at small $\omega$ for each system size that is
not seen for the other systems. Shown in Fig.~\ref{Lijin} are the results for the remaining systems
analyzed in Ref. \cite{Wang2021} (closed symbols). 


Figure \ref{Lijin} does not show any clear, systematic system size dependence of the exponent $\beta$. For the harmonic sphere system (HARM),
$\beta$ does appear to grow with system size, but the growth is within the uncertainty of the calculation. The two inverse-power-law
systems (IPL-10 and IPL-12) show no systematic change in $\beta$ with system size.

For the data presented in Fig.~\ref{Lijin} there is one outlier, and that is the IPL-10 system with 786 particles, see also Fig.~\ref{ipl10}.
We increased the number of glass samples for this system and system size from the 0.79 million samples used
in Ref.~\cite{Wang2021} to 2.4 million
samples. Shown in Fig.~\ref{N786} is
$I(\omega)/\omega^{4.5}$ for $N = 786$ with the improved statistics.  We can see that the upturn starts at lower $\omega$,
where the statistics are poor,
and there is a larger range of $\omega$ where $\beta \approx 3.5$ is a reasonable description of the data. We redid the
fitting for $\beta$ for this system  and the result is shown in Fig.~\ref{Lijin} as the open circle. With the improved statistics
$\beta$ is consistent with the other results.

\begin{figure}
\includegraphics[width=0.9\columnwidth]{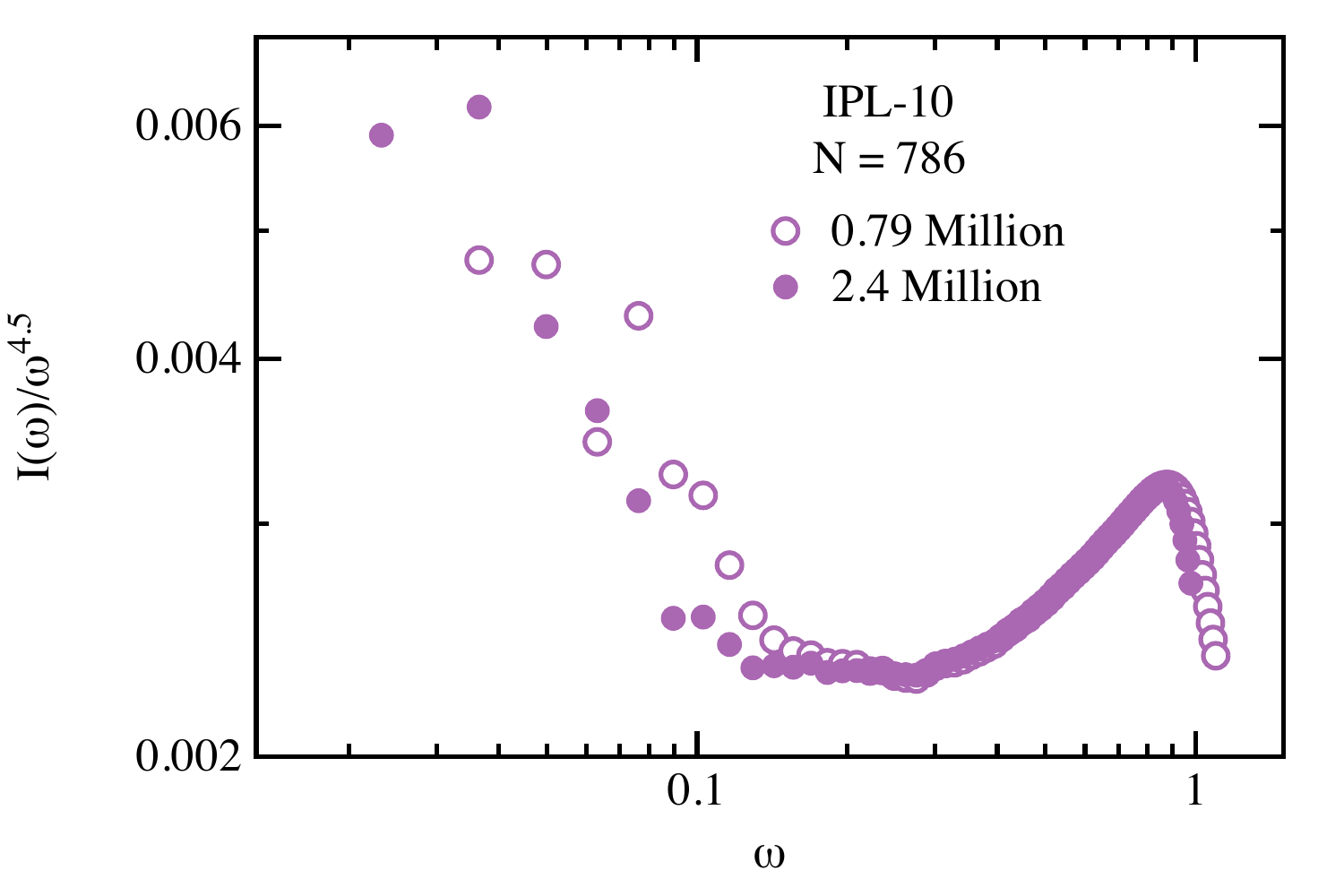}
\caption{\label{N786}Comparison of the scaled cumulative density of states for the IPL-10 system with N=786 for ensemble sizes
of 0.79 million configurations (open symbols) and 2.4 million configurations (closed symbols). }
\end{figure}

The change of $\beta$ with increasing the number of glass samples highlights the difficulty of obtaining an accurate value of
$\beta$ for two-dimensional systems. There are so few excess modes that many glass samples are needed to obtain
accurate results to do a direct calculation of the density of states. This was pointed out in Ref.~\cite{ Lerner2021jcp} where they utilized
non-linear mode analysis \cite{Kapteijns2020,Gartner2016} to analyze two-dimensional systems.


We find that all systems where $N > 100$ result in $\beta \approx 3.5$ for all our systems and we cannot
rule out $\beta \approx 3.5$ for $N \le 100$ with our data alone. We note that we do not believe
that one can obtain the scaling of the non-phononic density of states for $N < 100$.
We also find that these conclusions are consistent with the data presented in Ref.~\cite{Lerner2022}.
To argue for a finite size
effect the authors of Ref.~\cite{Lerner2022} fit the density of states, as opposed to the
cumulative density of states, and obtained different values
of $\beta$ for the power law
$D_{ex}(\omega) = A_{ex} \omega^{\beta}$. However, when the density of states is rescaled there is very convincing data collapse,
which suggests that all the results correspond to the same function and $\beta$ is the same for each system size.
To see convincing evidence of a system size dependence of $\beta$ one would need to see deviations from the
data collapse at low frequencies where we expect that the plane-wave-like modes do not contribute
to the density of states, and these deviations are not present in the density of states presented in Ref.~\cite{Lerner2022}.

We also don't find clear evidence for any system size dependence of $\beta$ in the cumulative density of states shown in
Fig.~2 of Ref.~\cite{Lerner2022}. We argued above that the scaling of the non-phononic density of states cannot be
inferred from systems of 100 particles or less. If we focus on the systems of greater than 100 particles we find that Fig. 2(d)
of Ref.~\cite{Lerner2022} shows that $\beta$ is the same for $N = 196$ and $N = 1600$ poorly annealed system.
According to Fig.~1(c) of Ref.~\cite{Lerner2022}
$\beta \approx 3.6$ for these systems. This value of $\beta$ is consistent with our results as well as the low frequency scaling
of the  cumulative density of states for the $N = 196$ and $N = 1600$ well-annealed system of Ref.~\cite{Lerner2022}.

All the systems discussed in this note and in Ref.~\cite{Lerner2022} are fairly small, and it is possible to study much larger systems.
The difficulty with larger systems is that the first plane-wave-like modes get pushed to lower frequencies and the frequency range where
only excess modes are expected to be found becomes
smaller. This difficulty is combined with the very limited number of non-plane-wave-like modes in two-dimensions.
We examined a $30\, 000$ particle system for the same IPL-10 model as used in Ref.~\cite{Wang2021} and found inconclusive results
 for the value of $\beta$ at the lowest frequencies examined, see Fig.~\ref{large}(b).
 These results were similar to those for three-dimensional systems in Ref.~\cite{Wang2022} in
 that $\beta$ was smaller than 3.5 at lower frequencies. This puzzling result
persisted despite having about 0.23 million configurations of $30\, 000$ particles. Examination
of these large systems deserves more work.

Non-linear excitation analysis has been used as a proxy to measure exponent $\beta$ in two-dimensional glasses
\cite{Gartner2016,Lerner2018,Lerner2021jcp,Kapteijns2020}.
Many properties of glasses are determined by the harmonic modes \cite{Mizuno2018,Wang2019,Flenner2019,Szamel2022},
and thus it is important to understand how the specific features of the harmonic modes influence these properties. We believe
it would be ideal to directly measure $\beta$ using harmonic modes, but find that this
is a very difficult task in two dimensions. The results presented
here strongly suggest that a harmonic modes-based calculation of $\beta$ for modes
below the first sound mode results in $\beta \approx 3.5$ rather than $\beta = 4.0$ for two-dimensional glasses.

Two-dimensional solids and liquids have characteristics different than their three-dimensional counterparts~\cite{Wang2021,Flenner2015}.
One characteristic for two-dimensional glasses is the very small density of excess modes, which makes
studying the excess modes difficult. There is evidence that the density of the excess modes scales as
$\omega^2$ for excess modes above $\omega_{T1}$ for two-dimensional systems \cite{Wang2021}.
Since $\omega_{T1} \rightarrow 0$
in the thermodynamic limit, it is unclear that the scaling of the modes below $\omega_{T1}$ is important
in the understanding of the behavior of two-dimensional glasses. Future work needs to focus on the influence of
excess harmonic modes on the low-temperature, thermodynamic properties of glasses.


We thank E. Lerner and E. Bouchbinder for useful discussions.  We also thank
E. Lerner and E. Bouchbinder for sharing some of their data published in Ref.~\cite{Lerner2022}, which we found to support the
conclusions presented here.  L. Wang acknowledges the support from  National Natural Science Foundation of China (Grant No. 12004001),
Anhui Projects (Grant Nos. S020218016 and  Z010118169), Hefei City (Grant No. Z020132009), and  Anhui University (start-up fund).
E.F. and G.S. acknowledge support from NSF Grant No. CHE-2154241.
We  also acknowledge Hefei Advanced Computing Center, Beijing Super Cloud Computing Center, and the High-Performance Computing Platform of Anhui University for providing computing resources.

\end{document}